# Study on the Availability Prediction of the Reconfigurable Networked Software System

Ling-zhong Meng[*], Min-yan Lu, Jing Lai, Xiao-jie Xu

*School of Reliability and System Engineering, Beihang University, Beijing, 100191, China*

This paper describes a multi-agent based availability prediction approach for the reconfigurable networked software system based on the complex system's model and simulation. Firstly, the definition of the features and availability of reconfigurable networked software system is introduced. Secondly, on the basis of the multi-agent based networked software system modeling method, an availability prediction method of the multi-agent based reconfigurable networked software system is proposed after the analysis of the reconfigurable behavior model. Finally, in order to verify the validity of this proposed method, an actual project is taken to predict its reliability. And the simulation result indicates that the availability of reconfigurable networked software system is significantly improved.

## 1. INTRODUCTION

Networked Software System (NSS) is a kind of Internet-based online complex software system of which topology structure and activity can be evolutionary dynamically [1]. At present, the network-based software system applications and services have become an important pillar of national economy's sustainable development, social life and national safety, and with the rapid development of the network service, NSS is also widely used in various fields, such as on-line bank, e-government, ERP software, military C4I software and so on.

For some stand-alone software, in order to meet the needs of users the software needs to be manual repaired, or can recover by rebooting within a certain time once some parts fail. However, for the Networked Software System used in the critical areas, it will cause serious consequences if the software failure cannot be repaired in time.

Reconfigurable software system refers to system which can be reconfigured in structure when its modules fail, according to different design principles by utilizing the dynamic adjustment of the structure in order to complete the customers' requirements. It can improve the availability of NSS to apply reconfigurable technology to the network software system, so that the software system failures can be restored in a relatively short period of time. DESEREC is an integrated project of the sixth framework program of the European Union under the "Information Society Technologies" priority, strategic objective "Towards a global dependability and security framework"[2]. One of the research content is "Enable Services Dynamic Configuration" which is to establish a secure, trusted network and service platform.

NSS is constituted by a number of subsystem, the elements which compose of the subsystem call each other to complete the required tasks, while the software emerged from the overall system reliability, safety and so on. The traditional methods include the availability researches which based on the features and services[3,4], the evolution analysis of component-based model,, system reliability modeling and assessment methods based on Bayesian Networks [5,6,7] and component-based software reliability prediction, metrics [8] and analysis approach [9,10]. However, these traditional methods cannot be applied in the NSS reliability prediction due to its large scale, complex functional interaction, the complexity of the composition of the components and the reliability of its components will change with outside influence.

The American scholar Santa Fe Institute has proposed the concept of the Agent to meet the complex behaviors of the complex system on the macro level. The system is seen as composed of by several autonomous agents. The interactions between the agents are the source of the macro mode of the system. Through the establishment of the agent model, these relationships can better be understood and explained. Agent technology is also applied in the field of software engineering. Agent-oriented software engineering is considered an important means of supporting the development of complex software systems [11]. On this basis, the Mobile Agent-based Mobile Agent mobile system's reliability assessment study [12, 13, 14] and security research [15] have been developed by applying the agent to different software design and mobile agent's network services.

---

[*] Ling-zhong Meng, email: menglz@dse.buaa.edu.cn

In this article, using the multi-agent modeling and simulation methods, the reconfigurable NSS is regarded as a complex system to research its reliability and availability [16]. The structure of the paper is as follows. Section 2 reviews a brief summary of the basic concepts of agent and the main ideas of software availability. Section 3 is to study the multi-agent based reliability modeling and simulation method based the analysis of the reconfigurable NSS features. Section 4 shows experimental studies for verifying the proposed simulation with the combination of the actual reconfigurable NSS. Section 5 contains the conclusions.

## 2. CONCEPTS AND METHODS

The dependability is proposed firstly by French scholar Laprie as a collective term in the 1985. In his view, dependability is a nature of the computer system, making the services provided by the computer system has been considered to be trustworthy. It provides a conceptual framework and a clear pattern of development for building dependability computer systems [17]. Laprie made it to be clear that dependability of software contain availability, reliability, safety, confidentiality, integrity, and maintainability in 2004[13]. Software Reliability is the probability of failure-free software operation for a specified time in a specified environment.

Agents are the actors that are able to presence to meet the design goals in a particular environment with autonomy, flexibility operations [19]. Agent's structure includes the various abstract components of itself, the status and role of each abstract component in the running agent and the interaction of these abstract components. It currently includes three categories: response structure, thinking structure and hybrid structure [20].

Multi-agent Systems (Multi-Agent System, MAS) is a composition of two or more of the Agents, each Agent is an independent behavior entity, and encapsulates the state and behavior, and interacts with each other through the communication among them. MAS architecture can be divided into centralized, distributed and hybrid.

The basic idea of MAS is to consider many subsystems as the composition of multiple autonomous agents, and the interaction between the agents is the origin of the macroscopic properties of the system. Through the establishment of multi-agent model, these systems can be better understood and analyzed. The essential characteristic of multi-agent simulation method is to establish a conceptual model of the actual system from the multi-agent perspective. The process is as follows: 1) Analyze the actual system; 2) establish a conceptual model; 3) build the simulation model; 4) analyze of simulation results; 5) conclusion.

## 3. MULTI-AGENT BASED RECONFIGURABLE NSS AVAILABILITY STUDY

Firstly the concept of the characteristics of reconfigurable NSS and the availability of the system is analyzed, then the multi-agent reconfigurable NSS modeling method is studied which includes the modeling methods of the component agent and connector agent, multi-agent coordination and interaction modeling approaches and reconfigurable modeling method, the multi-agent based reconfigurable NSS availability simulation method is proposed. The process of the multi-agent based reconfigurable NSS modeling and simulation method is as illustrated in Fig.1.

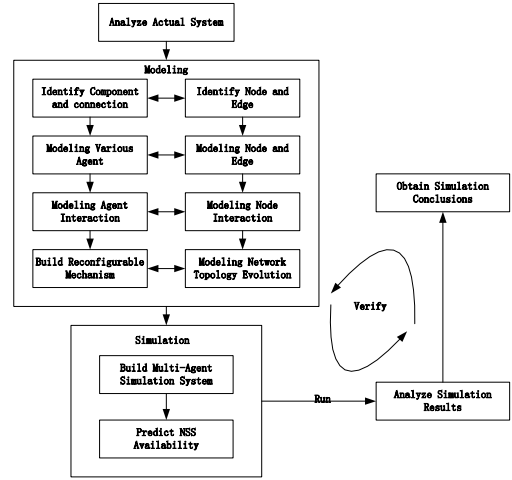

Fig. 1. Modeling and simulation framework based on MAS

### 3.1 Study of the reconfigurable NSS's Availability

The availability of reconfigurable NSS can be service-oriented, rather than the entire system-oriented. A reconfigurable NSS is able to ensure all services can be completed on schedule when its subsystem's component fail by the reconfiguration method, then the allocation strategy can be determined successful. For ease of presentation, the concepts used in the text are defined as follows:

**Definition 2.1 Software Service**

Software service refers to a set of software function provided by the NSS for the customers.

**Definition 2.2 Software Business**

The software business refers to the consumers' use of software services, including those aggregations of services involved in the software business, the services' usage mode and performance and quality requirements.

The reconfigurable NSS is a typical complex system, composed by a variety of software, and it has a very close relationship with the external environment and customers, its features are as flows:

1) Systemic. Reconfigurable NSS are closely related to the external environment, and reconfigurable NSS itself is an organic whole with a lot of interaction among subsystems. In general reconfigurable NSS is a multi-objective, multi-discipline and developmental system;

2) Evolvability. Under certain conditions, the reconfigurable NSS can be reconfigured with the occurrence of software failures to make the structure change, while the composition of software systems is constantly changing over time;

3) Random. Many random factors affect reconfigurable NSS, such as the spread of the virus, customers' impact, software defects, therefore reconfigurable NSS is a kind of random system;

4) Continuity. Reconfigurable NSS can provide continuous software business, management and maintenance of the database is online, generally will not stop.

The availability of the NSS measures in operational availability ($A_0$), refers to the total time of the software business provided by NSS, including operation time (OT), standby time (ST), the time of corrective maintenance (TCM), the time of preventive maintenance (TPM), the measurement model is as shown in (1).

$$A_0 = \frac{MUT}{MUT+MDT} \quad (1)$$

And MUT refers to the average working time, namely OT + ST, while the MDT is the average down time, namely TCM + TPM.

### 3.2. Multi-agent based reconfigurable NSS Modeling

Referring to software architecture, the composition of reconfigurable NSS are described with hierarchical models, object-oriented model and event-driven model. Hierarchical model is the basic structure of the entire software system simulation, object-oriented model aim to simulate the specific components in each layer, event-driven model used to simulate the interaction and communication among the components.

1. **The modeling research with the component agent and connector agent**

Based on software architecture analysis methods, software system can be considered as a combination of many components and connectors. In this mode, component is abstracted as a node, while connector is abstracted as a line. Therefore, the nodes and lines constitute the topology network. Multiple subsystems use connections to connect with each other, constituting the software topology network. Components and connections are abstracted as different structural types of agent, therefore nodes and lines of the topology network get more intelligent to promote the multi-agent availability prediction.

Component agent receives the required information from the network environment, make the information integration or information judgment according to the internal state, and then match the information in the Knowledge Base to formulate a planning, and finally generate a series of actions depending on the target, and thus have an impact on the external network environment. Therefore the component agent utilize the cognitive agent to accomplish the model, its formal description is as the following six-point groups.

Com_Agent :: = < Comi, Com_set, Com_KB, Com_PS, Com_GS, Com_AS >

Comi refers to the serial number of the component agent, Com_set refers to the set of states of the components of Agent, including component reliability, status and other information; Com_KB refers to the knowledge base of component Agent; the Com_PS refers to the component planning a collection of the Agent; Com_GS refers to the target set of component Agent; Com_AS refers to the action collection of component Agent.

Connector agents receives the required information from the network environment, then makes the information integration, and react with the internal rule library, and finally process information back to the network environment. Therefore connector agent utilize the reactive agent to accomplish the model, its formal description is as the following four-point groups.

Con_Agent :: = < Coni, Con_set, Con_PS, Con_AS >

Coni refers to the connector agent's serial number; Con_set refers to the state collection of the connector agent, including the connection reliability, status and other information; the Con_PS refers to the connector agent's planning collection; Con_AS refers to the connector agent's action set.

2. **Research on multi-agent based coordination and interaction modeling**

NSS consists of multiple subsystems. Each subsystem is composed by several component agent and connector agent. Subsystems connect with each other through different connection agent to transfer the data and control. A number of software services will be called in order to complete the user's software business, while each service involve several component agent and connector agent, namely, the simulation of each user's business use requires multiple component agent and connector agent. Therefore, the modeling and control of multi-agent coordination and interaction become very important. Based on this, a model which is suitable for NSS's multi-agent coordination and interaction is proposed here, shown in Fig. 2.

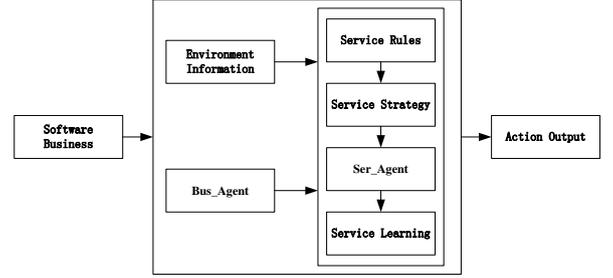

Fig. 2. Multi-agent coordination and interaction model

The proposed Bus_Agent and Ser_Agent is used to run a multi-agent coordination and interaction model, Bus_Agent refers to the agent which manages multiple services' use, the Ser_Agent set up the users' service of NSS, their formal description is as follows:

Bus_Agent :: =< Busi, Bus_set, Bus_ser_set, Bus_state>

Busi refers to the serial number of the Bus_Agent; Bus_set refers to the composition of the Bus_Agent collection, including the business' profiles, the transition probability of branches; Bus_ser_set refers to the set of services of business management; Bus_state refers to the status of Bus_Agent.

Ser_Agent:: =< Seri, Ser_S, Ser_P, Ser_Per, Ser_Trans >

Seri refers to the serial number of the Ser_Agent; Ser_S refers to the set of states; Ser_P refers to the set of perception; Ser_Per presents the perceptual function of E → Ser_P, the state-mapping of the environment refers to the perceptual input, E is the set of the environment; Ser_Trans refers to the decision-makingfunction of Ser_P * Ser_S → Ser_S, changing with according to the perceptual input and the current state to match the actual service.

3. **Study on reconfigurable modeling**

However, in the case of software failures occur, not all of the component agents and connector agents can be reconfigured to meet the user's requirement of the software business. During the design, the critical software business generally will be designed reconfigurable. The formal modeling of multi-agent reconfigurable process is as follows:

Re_Model :: = < Re_seti, <Re_act_set>, Re_PS, Re_Cor >

Re_seti refers to the serial number of the multiple agent set which participates in reconfigurable process; <Re_act_set> refers to the collection of each agent's action, in a reconfigurable process, each agent's effective actions constitute the set of actions; Re _PS refers to the service rule library involved in reconfiguration; Re_Cor refers to the service strategies, including coordination, collaboration and consultation between the multiple agent.

Re_act_set:: =< Re_act_seti, <Re_act _com>, <Re_act _con>>

Re_act _com::= <Fcomi, Rcomi_set >

Re_act _con:: = <Fconi, Rconi_set >

The Re_act_seti refers to the serial number of the reconfigurable action; <Re_act_com> refers to the set of reconfigurable component agent; <Re_act_con> refers to the set of reconfigurable connector agent, Fcomi means the serial number of the failed component agent; Rcomi_set refers to a group of component agent which can complete the Fcomi function of the failed component agent through reconfigurable actions; Fconi means the serial number of the failed connector agent; Rconi_set refers to a group of connector agent which can complete the Fconi function of the failed connector agent through reconfigurable actions.

## 3.3. Study on the availability simulation of multi-agent based reconfigurable NSS

There are some other functional agents in the availability simulation of the reconfigurable NSS except for component agent, connector agent, business agent (Bus_Agent), service agent (Ser_Agent) or reconfigurable agent. These agents are used to implement specific functions to ensure the smooth of the simulation operations and the collection of simulation data. These agents include Schedule Agent (Sch_Agent), Monitor Agent (Mon_Agent), Corresponce Agent (Cor_Agent), Data Agent (Dat_Agent) and GUI Agent (Gui_Agent), etc. Schedule Agent is to set the simulation time, decide when the simulation system starts and terminates the simulation when the conditions are met simulation; Monitor Agent collected the data during the simulation in order to achieve real-time display; Corresponce Agent used to provide multiple means of communication among the agents; Data Agent can handle various data during the simulation process to describe the system's macro performance; GUI Agent is used to receive input data and send output data. The simulation architecture is shown in Fig. 3.

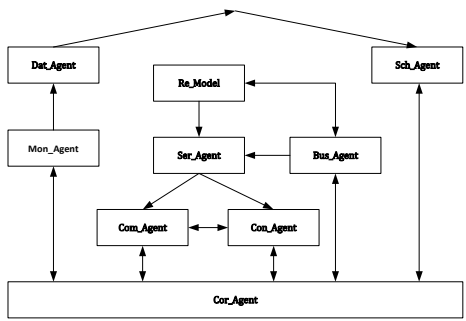

Fig. 3. Reconfigurable NSS simulation architecture

Various types of agent's properties and behavior models are described as follows:

Sch_Agent=<Schi, Sch_clock, Sch_Cori>, Schi is for the sequence number of the schedule agent; Sch_clock is for the simulation clock; Sch_Cori is for the connect-out Corresponce agent.

Mon_Agent=<Moni, Mon_type, Mon_Cori >, Moni is for the sequence number of the monitor agent; Mon_type is for the type of the monitor agent; Mon_Cori is for the connect-out corresponce agent.

Cor_Agent=<Cori, Cor_type, in_agents, out_agents>, Cori is for the sequence number of the Corresponce Agent; Cor_type is for the type of the Corresponce Agent, includes monitor agent corresponce, component agent corresponce, connect Agent corresponce; in_agents is for the connect-in agent; out_agents is for the connect-out agent.

Dat_Agent=<Dati, Data_type>, Dati is for the sequence number of data agent; Data_type is for the type of data agent,

Gui Agent=<Guii, Gui_type>, Guii is for the sequence number of the GUI agent; Gui_type is for the type of the GUI Agent.

In the simulation platform, a variety of agents' behavior is scheduled to coordinate by schedule agent in order to ensure the normal operations of the simulation system. The simulation clock utilizes the equal step method to promote service agent, connector agent, component agent, and ensure the time synchronization of each agent and update the status at the same time. The statistical time of different kinds of time in the availability metrics model is arrived through simulation clock, thus the prediction of the reconfigurable NSS availability will be obtained.

## 4. APPLICATION

An information processing system is selected to study its availability simulation here. The system is responsible for receiving all kinds of data to meet a variety of requirements of management and application. Meanwhile, the system output different kinds of special information for other systems in order to meet the diverse requirements of the users. Thus, this demand the systems must have very high availability. This information processing system is taken as an application case to study the availability prediction based on the multi-agent modeling and simulation method though designing it as a reconfigurable system.

As stated above, this proposed application is study on the simulation platform 'Netlogo'. Netlogo is a modeling and simulation platform, which is programmable and able to simulate the evolution of complex object behavior. It's especially suitable for the complex systems' modeling and simulation which is constantly changing with the interaction time.

In order to facilitate the system simulation, the following principles are set for the implementation of simulation.

1) Both component agent and connector agent have only two states, failure mode and normal running state;

2) The failure of the component agent and connector agent are independent with each other;

3) In order to accelerate the operation of the software system, the reliability of component agent and connector agent are both set to 0.9999;

4) Reconfigured behavior can be instantaneous with key business-related component failure occurs;

5) The using probability of each software business is the same.

On this basis, multi-agent method is utilized to make the availability model of the reconfigurable NSS. The simulation system has been run to get the final state which is shown in Fig. 4.

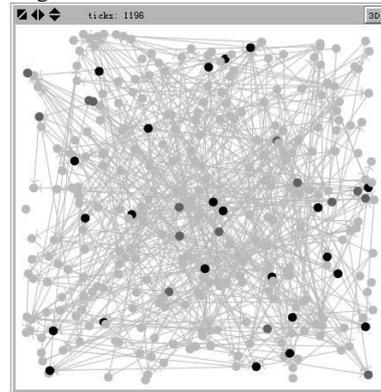

Fig. 4. Availability of simulation results

The nodes represent the component agent, amount to 306. Independent node refers to this component agent can provide some of the software business alone. Directed lines represent the connector Agent, amount to 459. Black nodes refers to the agent cannot be reconfigured to returned to normal once failure occurs, and these agents are the agent composed of non-critical software business. The dark gray nodes refers to the agent can be reconfigured to returned to normal once failure occurs, and these agents are the agent composed of critical software business.

The simulation system's availability curve is showed in Fig. 5 after running for some time. The horizontal axis represents the running time; the vertical axis represents the availability of the system, the value of availability metric interval is set to [50%, 100%] in order to display expediently. The black curve represents the availability variation tendency of the non-reconfigurable system; the gray curve represents the availability variation's tendency of the reconfigurable system.

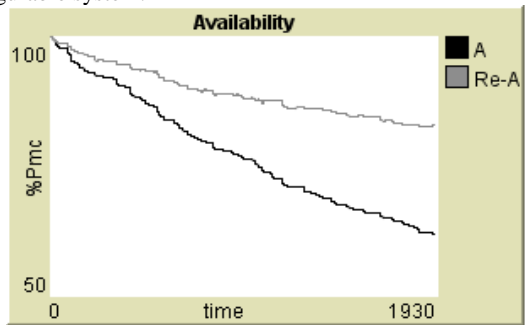

Fig. 5. Comparison to reconfigurable NSS and non-reconfigurable NSS

According to the analysis of the above results, the availability of the two systems is similar in the early running period of NSS. However, with the advance of simulation time, the reconfigurable system's availability is significantly higher than the availability of non-configuration system, and the high part is gradually increasing. This indicates that reconfigurable behavior has a significant role in the availability of NSS. Therefore, in some critical system, using the reconfigurable method can improve this kind systems' availability.

## 5. CONCLUSIONS

According to the multi-agent technology, this paper analyzes and solves the availability prediction of reconfigurable NSS from the perspective of complex systems. Multi-agent based modeling and simulation method is used to verify that the availability of reconfigurable system have obvious improvement than non-reconfigurable System with running time increasing.

In the future work, we can do further research from the following aspects: To Coordinate the number and costs of the reconfigurable component agent and connector agent with development costs; and based on works above, to research the selection criteria of the reconfigurable component agent and connector agent.